\documentclass[%
 reprint,
 prl,
 amsmath,amssymb,
 aps,
 superscriptaddress,
]{revtex4-2}

\usepackage{xcolor}
\usepackage{graphicx}%
\usepackage{bm}
\usepackage{amsthm}
\newtheorem{theorem}{Theorem}

\theoremstyle{remark}

\begin{document}

\title{Thermally-driven Quantum Reservoir Computing}

\author{Thiago Werlang}
\email{thiago.werlang80@gmail.com}
\affiliation{Instituto de F\'isica, Universidade Federal de Mato Grosso, Cuiab\'a, MT 78060-900, Brazil}

\author{Daniel Valente}
\email{valente.daniel@gmail.com}
\affiliation{Centro Brasileiro de Pesquisas F\'isicas, Rio de Janeiro, RJ 22290-180, Brazil}
\affiliation{Instituto de F\'isica, Universidade Federal de Mato Grosso, Cuiab\'a, MT 78060-900, Brazil}

\date{\today}

\begin{abstract}
Heat generation limits computation, classical and quantum, from single logic gates to entire data centers. However, an in principle limiting physical factor can sometimes be turned into a resource. In reservoir computing, energy dissipation is what gives a reservoir its fading memory. Here, we introduce a paradigm where the data injection itself is encoded on the time-dependent temperature of a thermal bath. We derive an exact expression for the nonlinear response of any fixed readout, and prove a rigorous memory--nonlinearity trade-off which turns out to be a structural necessity, rather than a tunable dynamical regime, since the same coupling that injects data also enforces equilibration. We illustrate both results with a driven qubit, using a single free parameter, and with a pair of coupled qubits, where the breaking of a discrete symmetry opens a fifteen-dimensional space to a single observable, yet that observable alone recovers just over half of it. This refutes the misconception that large Hilbert spaces guarantee computational power. Our results relocate the memory--nonlinearity trade-off from the reservoir's dynamics to its measurement channel, pointing to observable choice and bath engineering (instead of Hamiltonian control) as the physical levers for a thermally-native quantum reservoir computer.
\end{abstract}

\maketitle

\textit{Introduction}.-- Every physical implementation of computation is ultimately constrained by the thermodynamics of the substrate that carries it out. 
Reservoir computing (RC) delegates nonlinear processing to a fixed, high-dimensional transient dynamical system and confines all training to a linear readout~\cite{Maass2002,Jaeger2004,Verstraeten2007,LukoseviciusJaeger2009}, sidestepping the cost of training recurrent networks directly and enabling physical, non-digital implementations~\cite{Tanaka2019,Nakajima2020,Appeltant2011}.
Quantum reservoir computing (QRC) extends this to exploit the exponentially large Hilbert space and native many-body dynamics of quantum systems~\cite{Fujii2017,Nakajima2019}, where the dissipation of an open system, rather than an obstruction, supplies the fading-memory timescale an autonomous reservoir requires~\cite{Mujal2021,ChenNurdin2019,Sannia2024}. 
Heat's role in this literature, however, has remained passive. 
It sets a decay rate, or, in thermodynamic computing, supplies randomness~\cite{Conte2019,Camsari2019,Melanson2025}, but never itself carries the signal being processed.
Conventional QRC injects data coherently, either by writing each input into the state of an input qubit~\cite{Fujii2017,Nakajima2019} or by modulating the Hamiltonian~\cite{Govia2021}, and quantifies capacity as a property of the full internal state, typically characterized empirically, treating the reservoir as a black box probed through generic benchmarks~\cite{Fujii2017,Nakajima2019,MartinezPena2023}.

Here we introduce a distinct paradigm, \emph{thermally-driven} QRC, in which data are encoded through time modulation of a dissipative bath's temperature. 
Rather than benchmarking the reservoir as an opaque box, we take the opposite approach and open it, deriving what a fixed readout can extract directly from the algebra of the Hamiltonian, the dissipator, and the observables themselves. 
This matters because the dynamical expressiveness present in a reservoir is routinely suppressed by the projection intrinsic to any measurement channel~\cite{CindrakKrylov2025}. 
The operative bottleneck is not what the reservoir processes internally, but what a given observable can extract, and answering that question requires the reservoir's actual physics rather than a black-box characterization of its
outputs.

We make this precise by deriving an exact nonlinear-response expansion for a generic linear readout of a multi-level system coupled to modulated baths, and prove that it exposes a memory--nonlinearity trade-off imposed on the readout channel for \emph{any} fixed observable.
Such an antagonism is already documented as an emergent, dynamical property of the reservoir state, via the information-processing-capacity sum rule in classical RC~\cite{Dambre2012}, and in QRC, by a recent analysis that traces several seemingly unrelated frameworks to the same trade-off~\cite{Cindrak2026}.
We show instead that it is a structural consequence of the open-system generator algebra, not of any tuned dynamical regime.
The mechanism is direct. 
Because bath-mediated injection is simultaneously the excitation and relaxation channel, the coupling that generates higher-order nonlinearity inexorably suppresses cross-time memory, upstream of any measurement and distinct from
concurrent thermodynamic treatments~\cite{DingQiu2026}. 
The energy-efficiency motivation for neuromorphic substrates~\cite{Markovic2020} thus finds, in the thermal channel itself, both the computational resource and its fundamental constraint~\cite{Landauer1961}.

\textit{Thermal injection and a generic readout}.-- We model the reservoir as an $N$-level open quantum system with time-independent Hamiltonian $\hat H_0$, coupled to an input dissipative channel and an arbitrary number of passive ones (collected into $\mathcal L_{\mathrm{pass}}$). 
The density matrix obeys the Lindblad (GKSL) master equation~\cite{Lindblad1976,breuer2002theory},
\begin{equation}
\dot{\hat\rho}(t)=-\tfrac{i}{\hbar}[\hat H_0,\hat\rho(t)]
+\mathcal L_{\mathrm{in}}(\hat\rho(t))+\mathcal L_{\mathrm{pass}}(\hat\rho(t)).
\label{eq:gksl}
\end{equation}
Rather than writing data into the state or the Hamiltonian, the routes of conventional QRC~\cite{Fujii2017,Govia2021}, we inject data exclusively through the input channel’s bath temperature.
Engineering a bath with a controllable effective temperature is not merely a theoretical device. 
Superconducting-circuit experiments already use microwave waveguides carrying synthesized thermal noise as engineered baths, both to run a noise-powered quantum refrigerator~\cite{Sundelin2026} and to reset a qubit autonomously through a thermal gradient~\cite{Aamir2025}.
Each datum $x_k$ modulates the bath over a clock period $\tau$, entering $\mathcal L_{\mathrm{in}}$ in Eq.~\eqref{eq:gksl} through an affine shift of the input bath's occupation number on each dissipative channel $\omega$, $n(\omega,t)=n_0(\omega)+x(t)\,\chi(\omega)$, with $n_0$ the passive background occupation and $\chi$ a static spectral mask \cite{SM}.

The readout is left deliberately open. 
For any fixed Hermitian operator $\hat O$, define $J(t)=\mathrm{Tr}[\hat O(\hat\rho(t)-\hat\rho_{ss})]$, with $\hat\rho_{ss}$ the steady state of Eq.~\eqref{eq:gksl} at zero drive ($x=0$), which to linear order in the deviation $\Delta\vec\rho(t)=\vec\rho(t)-\vec\rho_{ss}$ (the state written as a vector in a fixed operator basis) is a fixed linear functional of the state, 
\begin{equation}
J(t)=\vec{\mathbf c}_R^{\,T}\Delta\vec\rho(t),
\label{eq:dJR}
\end{equation}
with $\vec{\mathbf c}_R$ the Liouville-space image of $\hat O$ \cite{SM}. 
Neither theorem below depends on which $\hat O$ is chosen, only on which reservoir directions are observationally inaccessible to it and the extent of the trade-off it exposes. 
We sample $J(t)$ at $V$ virtual nodes per cycle~\cite{Appeltant2011}, forming a feature vector that carries both $x_k$ and, through the inherited state, its history. 
A single linear regression then fits the output weights. 
This richness saturates once $V$ exceeds the reservoir's Krylov observability dimension $N_{\mathrm{eff}}$~\cite{CindrakKrylov2025}. 
This is generally much smaller than the reservoir's full state-space dimension $D=N^2-1$, and the gap between the two matters throughout what follows. 
A large Hilbert space bounds what is available in principle, but it is a fixed readout's own algebra with $\hat H_0$, not $N$ alone, that sets what can actually be extracted. 
Treating Hilbert-space dimension itself as a proxy for computational power is a recurring point of confusion in reservoir computing. 
The two reservoirs explicitly carried out by the end of this letter instantiate both extremes of the gap between $D$ and $N_{\mathrm{eff}}$ concretely.

Vectorizing in the $D=N^2-1$-dimensional Liouville space, $\Delta\vec\rho$ obeys \cite{SM}:
\begin{equation}
\Delta\dot{\vec\rho}(t)=\mathbf M_0\Delta\vec\rho(t)
+x(t)[\mathbf M_L\Delta\vec\rho(t)+\vec{\mathbf B}_{\mathrm{eff}}],
\label{eq:afim}
\end{equation}
a \emph{bilinear} system. 
Here $x(t)$ modulates the evolution operator itself, not merely a source term. 
$\mathbf M_0$ is the reservoir's free dynamics.
$\mathbf M_L$ and $\vec{\mathbf B}_{\mathrm{eff}}$, both originating in the same bath coupling, describe how the drive perturbs it. 
Stability requires $\mathbf M_0$ Hurwitz, with spectral gap $\mu_N=\min_a|\mathrm{Re}\,\lambda_a(\mathbf M_0)|>0$ setting the relaxation rate $1/\mu_N$, the reservoir's memory horizon. 
Computational nonlinearity requires $[\mathbf M_0,\mathbf M_L]\neq0$, since the injection must rotate the state out of $\mathbf M_0$'s eigenspaces rather than merely rescale it, a condition we now make precise for any fixed readout $\vec{\mathbf c}_R$.

\textit{Nonlinear response representation}.-- What follows characterizes not the reservoir's state, but what a single fixed readout of it sees, order by order in the drive.
\begin{theorem}[Nonlinear response representation]
\label{thm:volterra-main}
Let $\mathbf{M}_0$ be Hurwitz and diagonalizable, with the propagator constant $K_N$ \cite{SM}, and let $\vec{\mathbf c}_R$ be the readout covector of any fixed observable $\hat O$ (Eq.~\eqref{eq:dJR}). 
Within the convergence radius $\|x\|_\infty<\mu_N/(K_N\|\mathbf{M}_L\|)$, the transient readout is an exact nonlinear-response expansion in the drive, 
\begin{equation}
J(t)=\sum_{m=1}^{\infty}\int_0^t\!ds_1\cdots\!\int_{s_{m-1}}^{t}\!ds_m\;
\Gamma^{(m)}_{s_1,\dots,s_m}\prod_{l=1}^{m}x(t-s_l),
\label{eq:volterra}
\end{equation}
with order-$m$ response kernel
\begin{equation}
\Gamma^{(m)}_{s_1,\dots,s_m}=\vec{\mathbf c}_R^{\,T}
e^{\mathbf{M}_0 s_1}\mathbf{M}_L\,e^{\mathbf{M}_0\delta_2}\cdots
\mathbf{M}_L\,e^{\mathbf{M}_0\delta_m}\vec{\mathbf B}_{\mathrm{eff}},
\label{eq:kernel}
\end{equation}
where $\delta_j=s_j-s_{j-1}$. Proof in the Supplemental
Material~\cite{SM}.
\end{theorem}
Equation~\eqref{eq:volterra} is a nonlinear-response representation of the readout in the classical sense. 
The $\Gamma^{(m)}$ are the order-$m$ response functions of $J(t)$ to the bath drive, the same objects a spectroscopist would call the nonlinear susceptibilities and a systems theorist the Volterra kernels~\cite{boydchua1985}. 
Reading the reservoir through this hierarchy is what lets the memory--nonlinearity trade-off below be resolved order by order, for any fixed $\vec{\mathbf c}_R$.
Each kernel $\Gamma^{(m)}$ has a direct physical reading. 
It describes a cascade of $m$ elementary events: (1) the bath injects information into the reservoir at times $\{t-s_l\}$ (via $\vec{\mathbf B}_{\mathrm{eff}}$); (2) the perturbation freely propagates under the passive dynamics $e^{\mathbf{M}_0\delta_j}$ between successive interactions; and (3) at each intermediate step the modulation operator $\mathbf{M}_L$ mixes the accumulated response nonlinearly. 
The kernel thus encodes the correlations of the system at $m$ distinct past times simultaneously, a \emph{cross-time memory} structurally absent from linear reservoirs.

The first-order kernel ($m=1$) acts as a linear filter, preserving the injected frequencies. 
It alone sets the linear memory, and depends on neither $\mathbf{M}_L$ nor the higher orders. 
Kernels of order $m\ge2$ combine $m$ input modes and synthesize intermodulation frequencies of the drive harmonics, encoding the nonlinear content required by a task. 
If the operators commute, $[\mathbf{M}_0,\mathbf{M}_L]=0$, the free propagators merge, $e^{\mathbf{M}_0 s_1}\cdots e^{\mathbf{M}_0\delta_m}=e^{\mathbf{M}_0 s_m}$.
Therefore, the kernel loses all intermediate-delay dependence, collapsing to a static distortion with no cross-time correlations \cite{SM}, \emph{regardless of which observable $\vec{\mathbf c}_R$ reads it out}. 
The reservoir therefore acts as a \emph{parametric mixer followed by a spectral filter}. 
$\mathbf{M}_L$ generates new frequencies through its non-commutation with $\mathbf{M}_0$, and the eigenspectrum of $\mathbf{M}_0$ selects which of them reach the readout, any readout.

\textit{The dissipative trade-off}.-- A single inequality bounds the kernels of every order.
A memory--nonlinearity trade-off is well documented in reservoir computing, both classical~\cite{Dambre2012} and quantum~\cite{Cindrak2026}, but those results evaluate capacity as an emergent property of the reservoir's internal state under coherent (Hamiltonian) injection.
Whether an analogous bound survives thermal-bath injection, and projection onto an arbitrary fixed observable, is not obvious \emph{a priori}. 
The following theorem shows it does, as a hard constraint on the readout channel itself.

\begin{theorem}[Dissipative trade-off between memory and nonlinearity]
\label{thm:tradeoff}
For a thermally-driven reservoir obeying \eqref{eq:afim} with $\mathbf{M}_0$ Hurwitz and diagonalizable, and any fixed readout covector $\vec{\mathbf c}_R$, the order-$m$ response kernel satisfies
\begin{equation}
|\Gamma^{(m)}_{s_1,\dots,s_m}|\le
\mathcal{C}_m^{N}\,\|\mathbf{M}_L\|^{m-1}\,e^{-\mu_N s_m}.
\label{eq:tradeoff}
\end{equation}
Proof in the Supplemental Material~\cite{SM}.
\end{theorem}

Nonlinear expressivity of order $m$ grows as $\|\mathbf M_L\|^{m-1}$, while memory of depth $d$ decays as $e^{-\mu_N\tau d}$. 
Both $\mathbf M_L$ and $\mu_N$ are properties of the generator pair alone, independent of the readout. 
This is not a coincidence, since in thermally-driven encoding the injection channel is also the loss channel, so stronger injection necessarily accelerates relaxation \cite{SM}.
Both capacities are fed by the same unitary sector in different ways. 
Nonlinearity comes from the failure of this unitary sector to commute with $\mathbf M_L$, as shown above. 
Memory comes from the non-normality it creates in $\mathbf M_0$ together with the anisotropic damping, which lets a perturbation grow transiently before decaying~\cite{SM}.
The antagonism is therefore structural, fixing the optimal operating point at intermediate coupling rather than being tuned dynamically.

Our bound differs from the classical and quantum information-processing-capacity sum rules~\cite{Dambre2012,Cindrak2026}, which cap total capacity by $N_{\mathrm{eff}}$, the accessible dimension already introduced above. 
The difference is in both object and origin. 
Its object is the kernel of one fixed readout, bounded order by order, not a capacity of the reservoir state. 
Its origin is the coincidence of injection and loss channels in bath encoding, not the tuning of a dynamical regime.
It is also distinct from thermodynamic treatments that bound capacity against irreversible work~\cite{DingQiu2026}.

\textit{Results and Physical Discussion}.-- We now characterize the dissipative trade-off proved above in two archetypal quantum reservoirs. The first is a single qubit, and the second is a pair of coupled qubits. Both systems are analyzed with the same set of computational tasks. Our purpose is not to benchmark these reservoirs against state-of-the-art computational performance, but to exhibit the trade-off as a redistribution of computational capacity. Because of this, we use three degree-resolved capacities, each isolating one Volterra order and one memory character. The linear memory capacity, $\mathrm{LMC}=\sum_d C[x_{k-d}]$, probes the first-order kernel $\Gamma^{(1)}$ and represents the degree-one content (linear memory). The cross-time degree-two capacity $C_2^{\mathrm{cross}}=\sum_{d_1<d_2}C[x_{k-d_1}x_{k-d_2}]$ probes the intermediate-delay structure of $\Gamma^{(2)}$, the computational counterpart of the cross-time nonlinearity introduced above. The same-delay capacity $C_2^{\mathrm{inst}}=\sum_d C[x_{k-d}^2]$ isolates the instantaneous degree-two content, which requires no cross-time memory. All capacities are computed with the standard protocol of reservoir computing~\cite{Dambre2012}. We drive the reservoir with long sequences of independent inputs $\{x_k\}$, integrate the full master equation cycle by cycle with no expansion in the input, train a linear readout on one part of each sequence, and take each capacity as the coefficient of determination $R^2$ on a later part, averaged over independent sequences (see the Supplemental Material~\cite{SM}). In both reservoirs studied, each datum $x_k$ is carried by an AC carrier built from harmonics of $\omega_c=2\pi/\tau$,
\begin{equation}x(t)=x_k\sum_{i\ge1}a_i\cos(i\omega_c t+\varphi_i),\qquad t\in[k\tau,(k{+}1)\tau),\label{eq:encode}
\end{equation}
with no dc component and with the modulation depth capped by positivity of the thermal-bath occupation. Neither theorem above requires this specific form, since both hold for any bounded drive within their convergence radius.

\textit{Single-qubit reservoir}.-- The smallest reservoir is a single qubit with $\hat H_0=\tfrac{\omega_0}{2}\hat Z+\Delta\hat X$, read through the observable $\hat O=\hat Z$ and coupled to a single thermal bath with occupation $n_\omega(t)=n_{\omega,0}+x(t)$. For the Bloch vector $(\langle\hat Z\rangle,\langle\hat X\rangle,\langle\hat Y\rangle)$, the two generators of Eq.~\eqref{eq:afim} are
\begin{equation}
\mathbf M_0=\begin{pmatrix}-\kappa&0&2\Delta\\0&-\kappa/2&-\omega_0\\-2\Delta&\omega_0&-\kappa/2\end{pmatrix},\qquad
\mathbf M_L=-\gamma\begin{pmatrix}2&0&0\\0&1&0\\0&0&1\end{pmatrix},
\label{eq:tls-generators}
\end{equation}
with $\kappa=\gamma(2n_{\omega,0}+1)$~\cite{SM}. At $\Delta=0$ the population decouples from the coherences, and $\mathbf M_L$ commutes with $\mathbf M_0$. An input in a later cycle then only modulates the decay of the population left by an earlier one. With nothing to rotate the deviation between cycles, these modulations average out over each full period of the zero-mean carrier. Nonlinearity survives only within a single cycle, in $C[x_{k-d}^2]$, and the cross-time capacity $C[x_{k-d_1}x_{k-d_2}]$ vanishes at all orders. The $\hat Z$ readout then sees a single mode, $N_{\mathrm{eff}}=1$. A transverse field $\Delta\neq0$ couples population and coherences and makes the Bloch vector precess. The precession carries a deviation created at $t_1$ back and forth between population and coherences, so the effect of a later input at $t_2$ depends on how far the earlier deviation has rotated. This dependence is the cross-time memory that $C[x_{k-d_1}x_{k-d_2}]$ measures. Formally, $\Delta$ and the anisotropy of $\mathbf M_L$ together give $[\mathbf M_0,\mathbf M_L]\neq0$. The same precession lets $\hat Z$ reach all three Bloch components, so $N_{\mathrm{eff}}=D=3$.

The field $\Delta$ also sets how long the qubit remembers. The memory horizon is $1/\mu_N$, with $\mu_N$ the decay rate of the slowest mode of $\mathbf M_0$. At $\Delta=0$ the slowest modes are the coherences, which decay at $\kappa/2$. Below $\Delta^*=\omega_0/\sqrt2$, a damped precessing pair remains the slowest mode, and increasing $\Delta$ mixes the faster-decaying population into it, so $\mu_N$ grows and memory shortens. Above $\Delta^*$, the slowest mode is purely relaxational, and increasing $\Delta$ mixes the slower coherences into it, so $\mu_N$ falls again. A single parameter thus moves the qubit along the trade-off of Theorem~\ref{thm:tradeoff} non-monotonically~\cite{SM}.

\begin{figure}[!tb]
\centering
\includegraphics[width=0.99\linewidth]{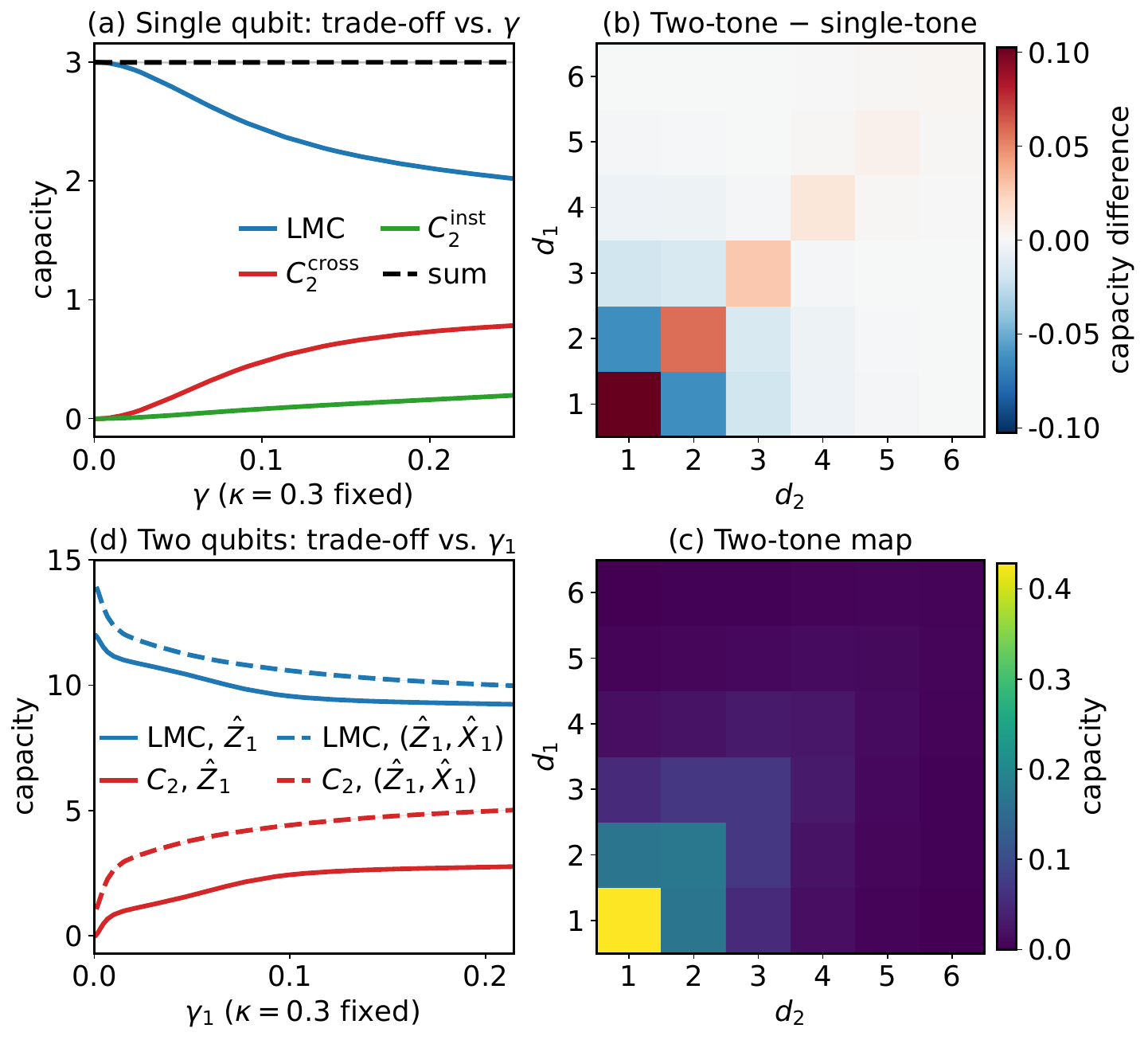}
\caption{The memory--nonlinearity trade-off of Theorem~\ref{thm:tradeoff} in both reservoirs, with the input amplitude fixed at $A=0.1$ ($x_k$ uniform in $[-A,A]$) in all panels.
Panel (a): single-qubit trade-off, swept in $\gamma$ at fixed $\Delta=0.4,\kappa=0.3$ (via $n_{\omega,0}=(\kappa/\gamma{-}1)/2$, single tone).
The sweep ends at the positivity limit $\gamma=0.25$.
$\mathrm{LMC}$ falls monotonically as $\gamma$ rises, $C_2^{\mathrm{cross}}$ and $C_2^{\mathrm{inst}}$ rise correspondingly, and the sum of capacities (dashed line) equals the rank $3$ within statistical error.
Panel (d): coupled-qubit reservoir with $\omega_z=1,\omega_x=0.7,J=0.5$ and identical baths, swept in $\gamma_1$ at fixed $\kappa=0.3$ (via $n_0=(\kappa/\gamma_1{-}1)/2$, two tones, $V=12$) up to its positivity limit $\gamma_1=0.214$.
Single readout $\hat Z_1$ (solid lines, rank $12$) versus joint readout $(\hat Z_1,\hat X_1)$ (dashed lines, rank $15$), both with drive period $\tau=2.0414$.
Here $C_2=C_2^{\mathrm{cross}}+C_2^{\mathrm{inst}}$.
Panels (b,c): single-qubit degree-2 capacity maps at $\kappa=0.3,\gamma=0.15,n_{\omega,0}=0.5$ and $\Delta=1.0>\Delta^*$, resolved over the $(d_1,d_2)$ grid (diagonal $=C[x_{k-d}^2]$, off-diagonal $=C[x_{k-d_1}x_{k-d_2}]$), with the same $A$ for both encodings.
Panel (b): two tones minus single tone. The second harmonic raises the same-delay capacities at the shortest delays and lowers the adjacent cross-delay ones, with a net gain in degree-2 capacity.
Panel (c): the two-tone map itself, showing where this capacity is concentrated.}
\label{fig:tls-combined}
\end{figure}

Figure~\ref{fig:tls-combined}(a) varies $\gamma$ at fixed $\kappa$ by compensating with the bath temperature, $n_{\omega,0}=(\kappa/\gamma-1)/2$. Because $\mathbf M_0$ depends on $\gamma$ only through $\kappa$, the free dynamics is identical at every point of the sweep, with the same eigenmodes, the same memory horizon, and the same temporal shape of the response. Only the injection changes. $\mathbf M_L$ grows as $\gamma$, and $\vec{\mathbf B}_{\mathrm{eff}}$ grows as $\gamma^2$, since $\vec\rho_{ss}\propto\gamma/\kappa$. The order-$m$ response therefore scales as $\gamma(\gamma A)^m$. Capacities are unchanged by an overall rescaling of the features, so they depend on $\gamma$ and $A$ only through the product $\gamma A$, and raising $\gamma$ strengthens the second-order response relative to the first-order one at fixed memory. As $\gamma$ grows, $\mathrm{LMC}$ falls monotonically while $C_2^{\mathrm{cross}}$ and $C_2^{\mathrm{inst}}$ rise monotonically. The rank stays at $N_{\mathrm{eff}}=3$, and the three capacities sum to $3$ within statistical error. Linear memory is therefore not lost to faster relaxation, which does not change along the sweep. The readout sees the past only through the three components of the Bloch vector, so it can reconstruct at most three independent functions of the input history. As $\gamma$ grows, the second-order response imprints products of past inputs on these same three components, and linear memory gives way to them. This is the trade-off of Theorem~\ref{thm:tradeoff} in its cleanest form, with $\Vert\mathbf M_L\Vert$ growing at fixed $\mu_N$.

To resolve how degree-2 capacity is distributed across delays, we compare single-tone and two-tone encodings of the input at the same amplitude $A$, both tuned to the precession frequency, $\omega_c=\Omega=\sqrt{\omega_0^2+4\Delta^2}$. We evaluate both at $\Delta=1.0>\Delta^*$, $\gamma=0.15$, $\kappa=0.3$, and $n_{\omega,0}=0.5$, where the single-tone same-delay and cross-delay capacities are of comparable size, so the degree-2 map carries weight both on and off its diagonal. Figures~\ref{fig:tls-combined}(b,c) are functions of the delay pair $(d_1,d_2)$, measured in drive periods $\tau=2\pi/\Omega$. The diagonal entries give the same-delay capacity $C[x_{k-d}^2]$, and the off-diagonal entries give the cross-delay capacity $C[x_{k-d_1}x_{k-d_2}]$. Panel (c) shows the two-tone map. Its largest entry sits at $d_1=d_2=1$, and the capacity fades within a few cycles in both delays. Panel (b) shows the difference from the single-tone map, isolating what the second harmonic changes. The change has opposite signs on and off the diagonal. The second harmonic raises the same-delay capacities at the shortest delays and lowers the adjacent cross-delay ones, most strongly at $(d_1,d_2)=(1,2)$. Beyond $d\simeq4$ the two encodings nearly coincide. Summed over the map, the gain on the diagonal outweighs the loss off it, so the total degree-2 capacity grows. Because the rank stays at $3$ for both encodings and the capacities sum to the rank, this net gain is paid for by an equal loss of linear memory. The encoding therefore redistributes capacity within a fixed rank rather than adding to it.

\textit{Two coupled qubits}.-- The second reservoir is a pair of qubits, each with a single-qubit Hamiltonian of the form above, coupled by an Ising term, $\hat H_0=\omega_z(\hat Z_1+\hat Z_2)+\omega_x(\hat X_1+\hat X_2)+J\hat X_1\hat X_2$. Each qubit couples to its own local thermal bath, and the input modulates only bath 1. The transverse term $\omega_x\neq0$ plays the role of $\Delta$ in the single qubit and breaks parity symmetry. Exchange symmetry between the qubits survives in $\hat H_0$ and, for identical baths, is broken only by the injection, which acts on bath 1 alone~\cite{SM}. Breaking parity gives the readout access to a larger Krylov subspace than at $\omega_x=0$. With identical baths, $\kappa=\gamma_1(2n_0+1)$ sets the common decay rate, and Fig.~\ref{fig:tls-combined}(d) sweeps $\gamma_1$ at fixed $\kappa=0.3$, as for the single qubit. Unlike the single qubit, $\mathbf M_0$ is not invariant along this sweep, since the local dissipators also couple two-qubit correlations to single-qubit operators at rates proportional to $\gamma_1$. Both readouts are sampled at $V=12$ virtual nodes per cycle with the same drive period, so the solid and dashed curves differ only in what is measured. The single readout $\hat Z_1$ resolves rank $12$ at every point, and the joint readout $(\hat Z_1,\hat X_1)$ resolves rank $15$. For both, $\mathrm{LMC}$ falls and the total degree-2 capacity $C_2=C_2^{\mathrm{cross}}+C_2^{\mathrm{inst}}$ rises monotonically over the whole range, and the capacities sum to the resolved rank within statistical error. Because $\hat Z_1$ and $\hat X_1$ each reach the full $D=15$ space on their own~\cite{SM}, the rise in rank from $12$ to $15$ comes from the number of features alone, since the rank of a multiplexed readout cannot exceed $\min(kV,N_{\mathrm{eff}})$ and the single readout supplies only $V=12$ features, against $2V=24$ for the joint one.

\textit{Conclusions}.-- In summary, we have introduced thermally-driven quantum reservoir computing, in which data are encoded in the occupation of a dissipative bath. For any fixed readout, we derived an exact expansion of the reservoir's nonlinear response, whose Volterra kernels describe cascades of injection, free propagation, and mixing by the modulation operator $\mathbf M_L$. This representation shows that cross-time memory requires $[\mathbf M_0,\mathbf M_L]\neq0$, whatever observable is read. We also proved that every kernel obeys an upper bound that grows as $\Vert\mathbf M_L\Vert^{m-1}$ and decays with the spectral gap $\mu_N$. Because in thermal encoding the same rate that sets the strength of the injection also enters the relaxation, this trade-off follows from the structure of the generator rather than from a particular operating regime.

The two reservoirs studied make this result concrete. In the single qubit, the transverse field $\Delta$ breaks the commutation of $\mathbf M_0$ and $\mathbf M_L$, creates cross-time memory, and lets the $\hat Z$ readout reach all three Bloch components. As the injection strengthens at fixed relaxation, $\mathrm{LMC}$ falls while the degree-2 capacities rise, with their sum fixed at the rank $3$. Linear memory is not lost to faster relaxation but converted into nonlinearity. Changing the carrier spectrum redistributes degree-2 capacity across delays within the same sum. In the coupled pair, the transverse field $\omega_x$ breaks parity and opens the full fifteen-dimensional space to a single observable. The same trade-off appears, and the rank rises from $12$ to $15$ when a second observable doubles the number of features.

More generally, both theorems are statements about what a fixed readout sees, not about the internal state of the reservoir. Earlier trade-off results treat capacity as a property of the state and bound it by the dimension of the accessible space~\cite{Dambre2012,Cindrak2026}. Here the bound holds order by order for the kernels of any observable, so the trade-off is set at the readout layer itself, where computation is actually extracted. Designing a thermally-driven reservoir is therefore a matter of choosing the observable, the operating point, and the drive spectrum, not merely of enlarging the Hilbert space. In this picture the bath coupling, usually treated as a source of decoherence or of passive memory, becomes the computational resource itself.

\begin{acknowledgments}
D.V. was supported by CNPq Grants 402074/2023-8 and 408990/2025-2. T. W. was supported by CNPq Grant 402074/2023-8. T.W. used Claude Opus 5.0 (Anthropic) to assist with the writing of this manuscript.
\end{acknowledgments}

\bibliography{references}

\end{document}